\documentclass[prl,twocolumn,showpacs,preprintnumbers,superscriptaddress,amsmath,amssymb,floatfix]{revtex4-1}

\usepackage{graphicx}

\begin{document}

\title{Calculating electron momentum densities and Compton profiles using the linear tetrahedron method}

\author{D.~Ernsting}
\email{d.ernsting@bristol.ac.uk}
\affiliation{H.H. Wills Physics Laboratory, University of Bristol, Tyndall Avenue, Bristol, BS8 1TL, United Kingdom}
\author{D.~Billington}
\affiliation{H.H. Wills Physics Laboratory, University of Bristol, Tyndall Avenue, Bristol, BS8 1TL, United Kingdom}
\author{T.D.~Haynes}
\affiliation{H.H. Wills Physics Laboratory, University of Bristol, Tyndall Avenue, Bristol, BS8 1TL, United Kingdom}
\author{T.E.~Millichamp}
\affiliation{H.H. Wills Physics Laboratory, University of Bristol, Tyndall Avenue, Bristol, BS8 1TL, United Kingdom}
\author{J.W.~Taylor}
\affiliation{DMSC - European Spallation Source, Universitetsparken 1, Copenhagen 2100, Denmark}
\affiliation{ISIS Facility, Rutherford Appleton Laboratory, Chilton, Oxfordshire OX11 0QX, United Kingdom}
\author{J.A.~Duffy}
\affiliation{Department of Physics, University of Warwick, Coventry, CV4 7AL, United Kingdom}
\author{S.R.~Giblin}
\affiliation{School of Physics and Astronomy, Cardiff University, Queen's Building, The Parade, Cardiff,
CF24 3AA, United Kingdom}
\author{J.K.~Dewhurst}
\affiliation{Max-Planck-Institut f\"{u}r Mikrostrukturphysik, Weinberg 2, D-06120 Halle, Germany}
\author{S.B.~Dugdale.}
\affiliation{H.H. Wills Physics Laboratory, University of Bristol, Tyndall Avenue, Bristol, BS8 1TL, United Kingdom}
\date{\today}

\begin{abstract}
A method for computing electron momentum densities and Compton profiles from \textit{ab initio} calculations
is presented. Reciprocal space is divided into optimally-shaped tetrahedra for interpolation, and the linear
tetrahedron method is used to obtain the momentum density and its projections such as Compton profiles. Results
are presented and evaluated against experimental data for Be, Cu, Ni, Fe$_3$Pt, and
YBa$_{2}$Cu$_{4}$O$_{8}$, demonstrating the accuracy of our method in a wide variety of crystal structures.
\end{abstract}

\maketitle

\section{Introduction}

Over the last few decades, studies of the electron momentum density (EMD) distribution have allowed insight into the
electronic structure of a wide variety of metallic, magnetic, and insulating
systems \cite{DUGDALE2014,KOIZUMI2011,DIXON1998,TAYLOR2002,DUFFY2010,SHUKLA1999,BARB2009,SAKURAI2011,KOIZUMI2001,BARB1999}.
The EMD is directly related to the square modulus of the Fourier transform of the electronic wave function, and may be written as
\begin{equation}
\begin{aligned}
 \label{emd}
 \rho({\bf p}) = \sum_{{\bf k},j} n_{{\bf k},j} \left| \int \psi_{{\bf k},j}({\bf r}) {\rm{e}}^{-{\rm{i}}{\bf p} \cdot {\bf r}}~{\mathrm d}{\bf r} \right|^2, 
\end{aligned}
\end{equation}
where $\psi_{{\bf k},j}({\bf r})$ is the wave function of an electron with the wavevector ${\bf k}$ in band $j$,
and $n_{{\bf k},j}$ is its occupation. As the EMD only contains contributions from occupied states, its measurement
allows the Fermi surface (FS) topology of metals to be determined \cite{DUGDALE2014}. Moreover, the close
relationship of the EMD to the electronic wave function has been exploited to great effect in studying wave function
coherence, bonding, and orbital character in a variety of complex materials
\cite{WEYRICH1979,SHUKLA1999,BARB1999,KOIZUMI2001,BARB2009,SAKURAI2011}. Significantly, comparisons between measured
and calculated momentum densities provide a rigorous means of testing, and refining theoretical models
\cite{LAMPLATZMAN1974,CARDWELL1989,KUBO1999,MAJOR2004}.

Inelastic x-ray scattering in the regime of very high energy transfer (so-called Compton scattering) is a powerful
experimental probe of the EMD \cite{LAVEROCK2007,KOIZUMI2011,SAKURAI2011,DUGDALE2006}.
In a Compton scattering
experiment, for a fixed momentum transfer, the projection of the bulk EMD along the scattering vector (taken as the
z-direction of a Cartesian coordinate system) can be extracted. In spin-polarised systems, the total EMD is comprised
of different contributions $\rho_{\uparrow}$ and $\rho_{\downarrow}$, from spin-up and spin-down electrons. Magnetic
Compton scattering \cite{DUFFY2013} measures the projection of the spin-resolved momentum density. Here components arising
from spin-paired electrons cancel, allowing the spin magnetic moment \cite{DUFFY2010}, and spin polarisation to be
determined \cite{UTFELD2009}. Compton profiles (CPs), $J(p_z)$, and magnetic Compton profiles (MCPs),
$J_{\rm{mag}}(p_z)$ are defined as 
\begin{equation}
 J(p_z) = \iint \left[\rho_{\uparrow}({\bf p})+\rho_{\downarrow}({\bf p})\right]~{\mathrm d}p_x{\mathrm d}p_y,
\end{equation}
and
\begin{equation}
 J_{\mathrm{mag}}(p_z) = \iint \left[\rho_{\uparrow}({\bf p})-\rho_{\downarrow}({\bf p})\right]~{\mathrm d}p_x{\mathrm d}p_y,
\end{equation}
respectively.

The momentum density (or spin density in momentum space) in two or three dimensions may be experimentally recovered
by measuring a series of profiles along different crystallographic directions, and employing a suitable tomographic
method \cite{KONTRYM-SZNAJD2005}. 

It is worth noting that the once-projected momentum density may also be probed via the two-dimensional angular
correlation of annihilation radiation (2D-ACAR) technique \cite{EF1995,DUGDALE2014}, which measures the electron-positron
momentum distribution (that is, the electron momentum density as seen by the positron). The inclusion of the positron
wave function, which is beyond the scope of the present work, can have a significant influence on the shape of the
momentum distribution \cite{RUSZ2005,LAVEROCK2010}.

Great insight into electronic structure is available in comparing measured momentum distributions 
with the results of density functional theory (DFT) calculations.
It should be emphasised that although the Kohn-Sham orbitals reproduce the
ground-state electron density in real space (for a given exchange-correlation functional), this is not true in
momentum space. However, as pointed out by Lam and Platzman \cite{LAMPLATZMAN1974}, 
a correction is possible. 
These so-called Lam-Platzman
corrections (which here, as in many other works, are not calculated) are isotropic, whereas 
discrepancies between calculations and experiment tend to be anisotropic
\cite{SHIOTANI1993,SAKURAI1999,HUOTARI2000}. We would also like to point out that the \textit{exact}
EMD (for a given exchange-correlation functional) can be calculated within
reduced density matrix functional theory (RDMFT) \cite{SHARMA2013} and the closely related natural orbital
functional theory (NOFT) \cite{GOEDECKER1998}, but such calculations tend to be very computationally
expensive compared to DFT. However, recent approximations to NOFT have been developed which make the
estimation of the occupation numbers in Eq. \ref{emd} for the exact EMD much more computationally efficient
\cite{BARB2014}.

Several methods exist for calculating the EMD within DFT, with those based upon the Korringa-Kohn-Rostoker
(KKR) \cite{BANSIL2001,EBERT2011,SPRKKR} and linearised muffin-tin orbital (LMTO) \cite{BARB2003} formalisms
being two of the most common, whilst other groups maintain their own in-house codes for calculating the EMD
within a full-potential DFT framework \cite{WAKOH2002,WAKOH2004,NAGAO2008,BROSS2012,PYLAK2013}. In these
methods, projections of the EMD are generally obtained by either evaluating the EMD on a unique grid for
each direction, or by calculating the EMD once on a certain grid, and interpolating the function on to a
suitable set of points for the desired integration. The latter method is often preferred as, once the momentum
density is evaluated on the initial grid the time taken to produce each profile is much less, with no
appreciable loss in accuracy as long as a suitable method is chosen for interpolation. 

Wakoh \textit{et al.} utilise the linear tetrahedron method
\cite{JEPSON1971,LEHMAN72,RATHFREEMAN1975,KAPRZYK1986,BLOCHL1994,KAPRZYK2012} to interpolate and project
the EMD along arbitrary directions. Their approach involves evaluating the EMD on a fixed grid of points
in reciprocal space, and then splitting these into a predetermined set of tetrahedra. Linear interpolation
is used to find a series of planes within each tetrahedron perpendicular to the scattering vector, and
Compton profiles are obtained as a sum over the approximate contributions from occupied planes
\cite{WAKOH2002,WAKOH2004}. Whilst this method provides results in good agreement with experimental
data \cite{WAKOH2002_Fe3Pt,TOKII2006}, it is only formulated for systems with cubic symmetry, and only
produces Compton profiles, rather than the once integrated (2D projection), or full 3D EMD. Also, as
the authors note, the result will vary depending upon the choice of tetrahedra \cite{WAKOH2004},
meaning that in general a single predetermined set will not give the most accurate interpolation.

Here, we build upon the work of Wakoh \textit{et al.} by presenting a method for obtaining the EMD
(and its projections) by the linear tetrahedron method, for \textit{any} crystal structure. The EMD
is obtained from the band energies and the plane wave expansion of the valence electron wave
functions of a converged DFT calculation at a set of $\bf{p}=\bf{k}+\bf{G}$ points, where {\bf G} is
a reciprocal lattice vector. We employ an algorithm to divide reciprocal space into optimally-shaped
 tetrahedra (for interpolation) before interpolating the EMD, ensuring accurate interpolation
irrespective of crystal structure. The code is written as a module for the popular, freely
available, and open source, full-potential linearised augmented plane wave (FP-LAPW) DFT code,
\textsc{ELK} \cite{ELK}, and can produce the full three-dimensional momentum density, or its
one or two-dimensional projections. A description of the technical implementation is provided
in the next section.

\section{Numerical method}

\subsection{Calculation of the momentum density within the irreducible Brillouin zone}
The plane wave expansion of the electron wave function in the irreducible Brillouin zone (IBZ) is calculated
by Fourier transforming the real space wave functions on to an extended set of $\bf{p}$-points in reciprocal
space, to describe the momentum density adequately (highly localised electrons may have momenta $> 16$ a.u.).
The momentum density at each point is then calculated for every state and spin via Eq. \ref{emd} (excluding
the occupation numbers from Eq. \ref{emd}, which are applied at the time of interpolation). The calculation
of the EMD in the irreducible wedge of momentum space only needs to be run once for a converged calculation.

\subsection{Splitting of reciprocal space into tetrahedra}
In \textsc{ELK}, the starting {\bf p}-point grid consists of a series of identical parallelepiped submesh cells
(SMCs) which tile to fill reciprocal space. The submesh cells are the same shape as the reciprocal unit cell of
the crystal, but are constructed from fractions of the reciprocal lattice vectors. For a non-cubic system, the
SMC may have relatively large distances between points on the original mesh. In order to interpolate the EMD as
accurately as possible, we split the submesh into tetrahedra where the distance between vertices is minimal. This
splitting is achieved using an algorithm based upon the Delaunay criterion \cite{DEL34}, implemented in three
dimensions. We provide a brief description of the method below (for a more thorough description, see Ref. \cite{BERG2008}).

The splitting algorithm creates a $root$ tetrahedron which contains one SMC. The points of the cell are incrementally
added, and the containing tetrahedron is split into new tetrahedra about each point.
The Delaunay criterion stipulates that if the circumscribed sphere of each tetrahedron does not contain a vertex
of any other, the minimum internal angle of the set of tetrahedra will be maximised (so that they will have small edge
lengths, and be optimally shaped for interpolation). As each new tetrahedron is created, it is checked to see if it satisfies
this rule and, if not, the rule is enforced through a transformation involving switching vertices with adjacent tetrahedra.
This check is performed recursively upon all transformed tetrahedra until the criterion is satisfied. When all the
points of the SMC have been added, the $root$ tetrahedron is removed, leaving the optimally-split SMC.

\subsection{Interpolation of the momentum density}
The occupation numbers for each {\bf k}-point and state are calculated from the energy eigenvalues using the
Fermi-Dirac distribution, in principle allowing for the calculation of a finite-temperature momentum
distribution. This also allows shifts of the relative Fermi energy for individual states to be performed
if required (sometimes small rigid energy shifts of the theoretical electronic band structure are necessary
to bring a calculated Fermi surface into agreement with experiment \cite{MAJOR2004,UTFELD2010}). The total
EMD is obtained as the occupation-weighted sum over the contributions from all states (but can, if required,
be resolved by state), and expanded from the irreducible part to the whole of momentum space for
interpolation using the point-group symmetries of the lattice.

To allow projection along a specific direction in momentum space (for example, the scattering vector of a Compton
scattering experiment), the EMD is interpolated onto a cubic mesh, which may be aligned along any arbitrary direction,
using the linear tetrahedron method within the optimally-shaped tetrahedra. 

The computational demands of the code depend sensitively upon the ${\bf k}$-point mesh of the original
calculation and the number of ${\bf p}$-points, but remain within the capabilities of modern desktop computers.
For a typical CP calculation of a single atom system such as Cu, over 511 ${\bf k}$-points, the code requires
around 1 Gb of memory, whereas for a more demanding five atom cubic system with a similar number of
${\bf k}$-points, the requirement approaches 4 Gb.
 
\section{Results}
In order to demonstrate the flexibility of our method, we have calculated the EMD of a broad variety of systems
with different crystal structures. Systems were chosen which have been the focus of both Compton scattering
experiments and theoretical studies, to allow comparison with both.

\begin{figure}
\centering
\includegraphics[width=1.0\linewidth]{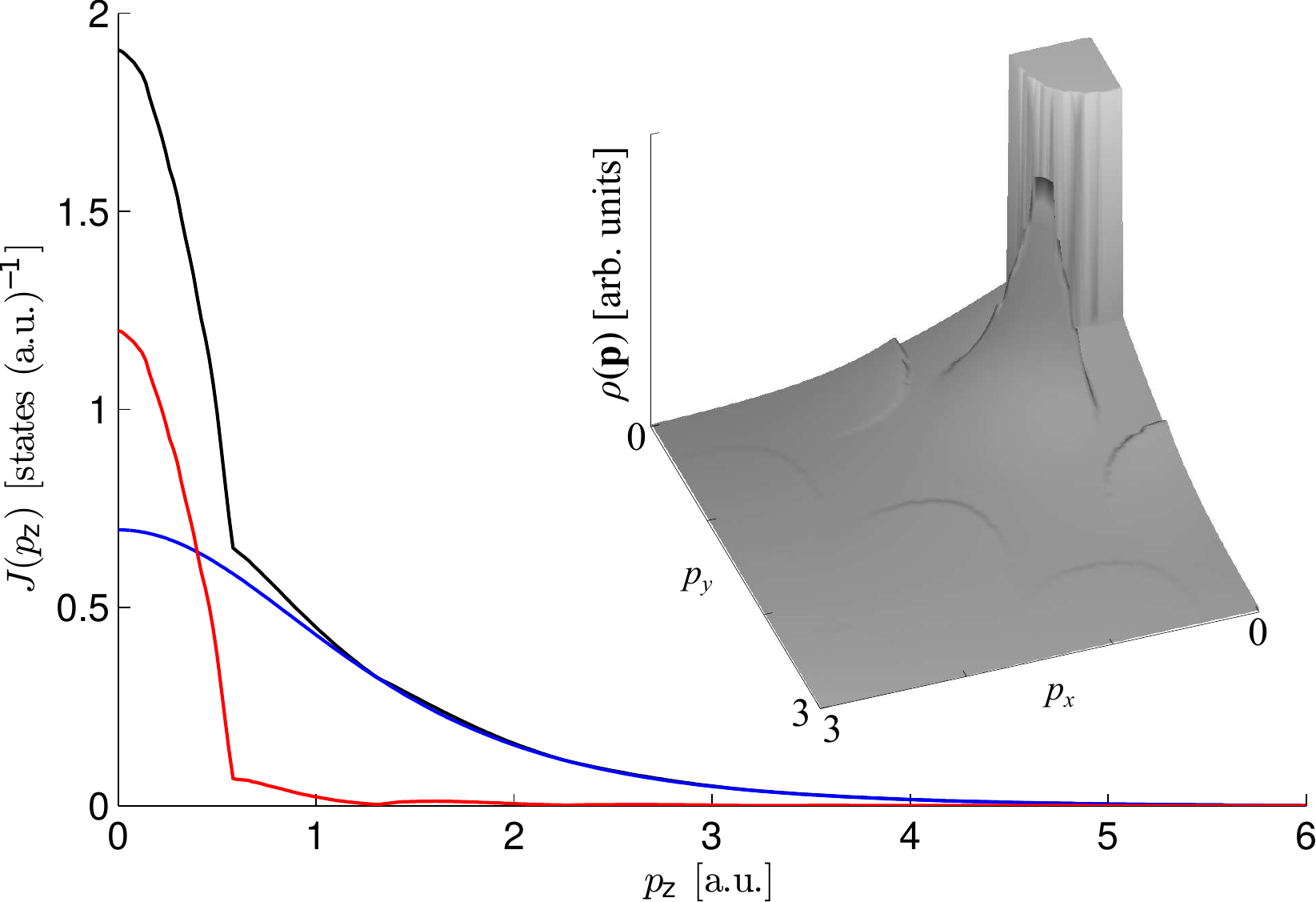}
\vspace*{-0.25in}
\caption{Calculated Compton profile of Li for a scattering vector along the [100] direction. The total profile
is shown (black line) as well as the separate contributions from the semi-core 1$s$ state (blue line), and
the 2$s$ valence (red line). Inset: (001) plane of the EMD, showing signatures of the Fermi surface.
The plot has been truncated in $\rho$ to highlight these anisotropic features.}
\label{li}
\end{figure}

As an example, the Compton profile of Li along the [100] direction is shown in Fig. \ref{li}. The shape
of the profile is a combination
of a broad isotropic Gaussian-like contribution from the more localised electrons (in this case the semi-core
1$s$ state), and a parabolic free-electron-like contribution from the valence electrons (the 2$s$ state).
Experimentally, the core contribution is often uninteresting (and unwanted), whilst the valence will
be anisotropic, and contain information about bonding or the Fermi surface. Indeed, the presence of the
Fermi surface is visible in the plane through the EMD of Li shown in the inset of Fig. \ref{li}.

To aid comparison with experimental and theoretical spectra from the literature, it is common to
calculate the directional difference between Compton profiles. This has the advantage of removing the isotropic
contribution to the momentum density from the core electrons (which tend not to be included in calculations of
the momentum density), whilst highlighting the anisotropic contribution from the valence electrons.
Another benefit of inspecting the anisotropic part of the EMD is that the isotropic Lam-Platzman correction
will have no effect (and therefore does not need to be calculated). 

All calculations were performed using the same exchange-correlation functional as the
calculations from the literature (where applicable), and have been convoluted with a Gaussian with a full width
at half maximum (FWHM) equal to the experimental resolution.

\subsection{Compton profiles of Be and Cu}
\begin{figure}
\centering
\vspace*{-0.06in}
\includegraphics[width=1.0\linewidth,trim=1.05in 0in 0.8in 0in, clip=true]{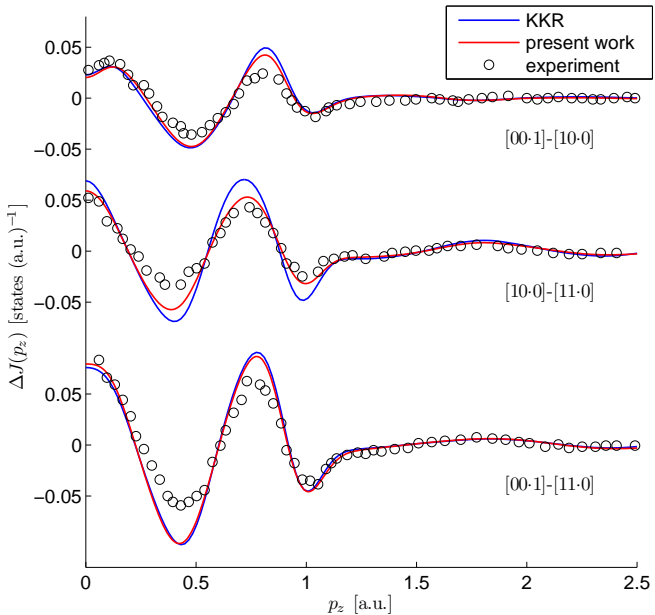}
\vspace*{-2.1in}
\caption{Directional differences of the $[10\cdot0]$, $[11\cdot0]$ and $[00\cdot1]$ Compton profiles of Be from
the experiments (open circles), and KKR calculations of Huotari \textit{et al.} (blue line) \cite{HUOTARI2000},
and those calculated by our method (red line). The authors of Ref. \cite{HUOTARI2000} note that the error on
their data is less than the symbol size.}
\label{be}
\end{figure}

The electronic structure of hcp Be was calculated using the lattice constants $a=4.3289$ a.u. and $c=6.7675$ a.u.,
with 1197 {\bf{k}}-points within the IBZ. The directional differences of Compton profiles calculated along
the $[10\cdot0]$, $[11\cdot0]$ and $[00\cdot1]$ directions are shown in Fig. \ref{be}, compared with the experimental
data of Huotari \textit{et al.} obtained at a momentum resolution of 0.16 a.u. \cite{HUOTARI2000}, and their KKR
calculations utilising the von Barth-Hedin local density approximation (vBH-LDA) to the exchange-correlation functional
\cite{VBH72}. The profiles calculated by our method agree very well with the experimental data, and show slightly
improved agreement compared to KKR calculations of Ref. \cite{HUOTARI2000}.

\begin{figure}
\centering
\vspace*{-0.06in}
\includegraphics[width=1.0\linewidth,trim=1.05in 0in 0.8in 0in, clip=true]{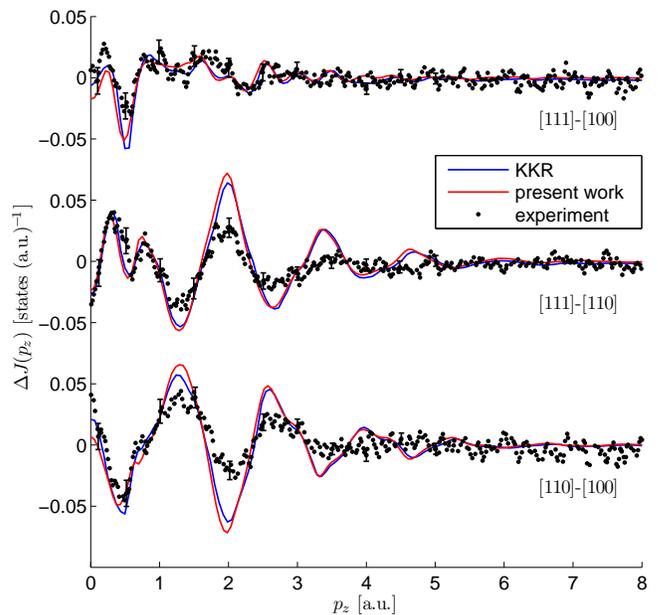}
\vspace*{-2.1in}
\caption{Directional differences of the Compton profiles of Cu for the [100], [110], and [111] directions
from the experiments of Sakurai \textit{et al.} (filled circles), their KKR calculations (blue line)
\cite{SAKURAI1999}, and our method (red line).}
\label{cu}
\end{figure}

The EMD of fcc Cu was calculated using a lattice constant of 6.8242 a.u., on 511 {\bf k}-points within the IBZ.
Figure \ref{cu} shows calculated anisotropies between the [100], [110], and [111] directions, compared to the
experimental data of Sakurai \textit{et al.}, which has an experimental resolution of 0.12 a.u., and their own
vBH-LDA KKR calculations \cite{SAKURAI1999}. Whilst the calculated profiles are in close agreement with the KKR
calculations of Sakurai \textit{et al.}, they do not agree quite as well with the experimental data, and both
calculations overestimate the directional anisotropy. This is a well known artifact in DFT calculations, which
has been attributed to the neglect of non-local electron correlation effects \cite{SHIOTANI1993}. This
disagreement has previously been reduced in Cu through the application of the so-called self-interaction
correction \cite{KUBO1999}.

\subsection{Spin resolved momentum densities of Ni and Fe$_{3}$Pt}

\begin{figure}
\centering
\vspace*{0.1in}
\includegraphics[width=1.0\linewidth]{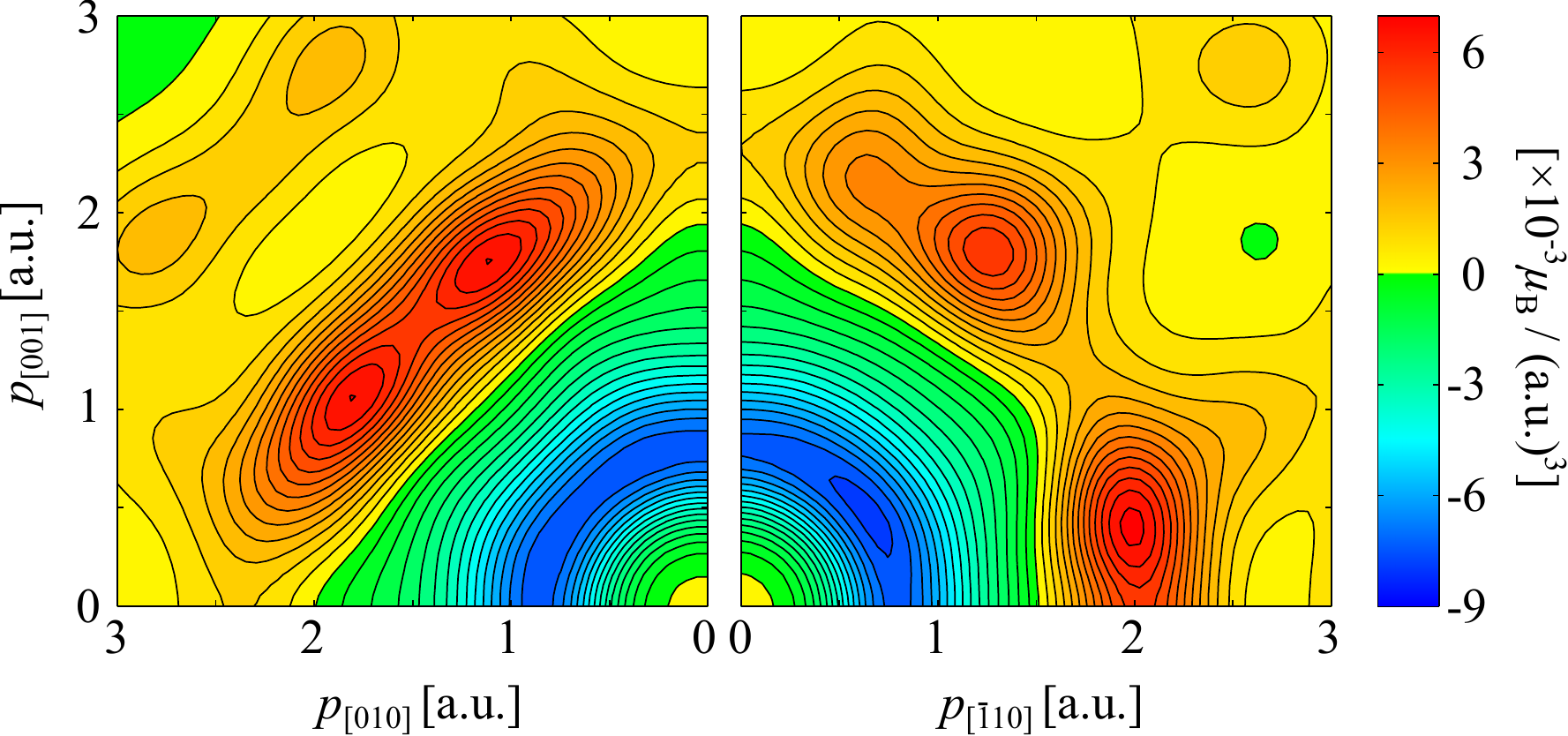}
\caption{Spin momentum density of the (100) (left), and (110) (right) planes of Ni. The contour interval is
set to $0.5\,\times10^{-3}\mu_{\rm{B}} \rm{/ (a.u.)^{3}}$.}
\label{ni2d}
\end{figure}

The electronic ground state of ferromagnetic fcc Ni was calculated with a lattice constant of 6.644 a.u.,
and using 511 {\bf{k}}-points in the IBZ. In order to compare our calculation with a recent 3D reconstruction
of the spin momentum density of Ni by Nagao \textit{et al.} \cite{NAGAO2008}, the 3D spin momentum density
was obtained on grids oriented along the cubic [100] and [110] directions, and convoluted with a 3D gaussian
of FWHM equal to 0.52 a.u. before normalisation to the calculated moment of 0.586~$\mu_{\rm{B}}$. The (100)
and (110) planes of the calculated distribution are shown in Fig. \ref{ni2d}, and agree very well with the
planes through the experimental data of Nagao \textit{et al.}, visible in parts (a) and (b) of Fig. 4 in
Ref. \cite{NAGAO2008}, respectively. There is a slight discrepancy at ${\bf p} = 0$ a.u., where the calculation
finds a higher spin density than is evident from experiment (also seen in the calculation in Ref. \cite{NAGAO2008}).
This discrepancy has previously been noted in MCPs calculated within both LMTO and full-potential
formalisms \cite{DIXON1998}, and can be rectified with small rigid shifts of the bands close to the Fermi
level \cite{MAJOR2004}. It would be interesting to see if this is also corrected in dynamical mean-field
theory calculations, which improve upon the agreement between experimental MCPs and those calculated within
the LDA \cite{BENEA2012,CHIONCEL2014}.

\begin{figure}
\centering
\vspace*{-0.2in}
\includegraphics[width=1.0\linewidth,trim=0.98in 0in 0.8in 0in, clip=true]{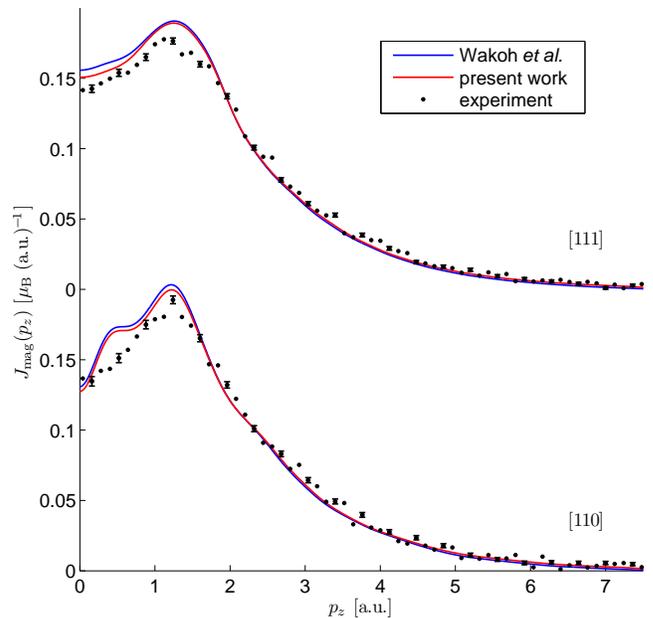}
\vspace*{-2.1in}
\caption{Magnetic Compton profiles of Fe$_{3}$Pt projected along the [111] and [110] directions. The results of
our method are shown (red line) with the experimental data of Taylor \textit{et al.} (filled circles)
\cite{TAYLOR2002}, and those of another computational study which utilises the linear tetrahedron method to
obtain MCPs from full-potential calculations (blue line) \cite{WAKOH2002_Fe3Pt}.}
\label{fe3pt}
\end{figure}

The electronic structure of ferromagnetic Fe$_{3}$Pt, which has a simple cubic structure, was calculated using
a lattice constant of 7.086 a.u., with 560 {\bf k}-points in the IBZ, using the Perdew-Burke-Ernzerhof
generalised gradient approximation (PBE-GGA), to the exchange-correlation functional \cite{PBE1996}. Calculated
MCPs for the [110] and [111] directions are shown in Fig. \ref{fe3pt}, compared with the experimental MCPs
of Taylor \textit{et al.} measured at 300~K with a resolution of 0.40 a.u. \cite{TAYLOR2002}, and
the calculations of Wakoh \textit{et al.}, obtained from full-potential DFT calculations by the linear tetrahedron
method \cite{WAKOH2002_Fe3Pt}. Our calculations have been normalised to the calculated moment of 8.54~$\mu_{\rm{B}}$.
Whilst both calculations reproduce the overall shape of the MCPs well, the result from our method is marginally
closer to the experiment in the low momentum region ($p_{z} <$ 1.5 a.u.).
As both calculations are produced by the linear tetrahedron method, the resulting MCPs should be effectively
the same, and as such the slight improvement at low momentum is possibly due to the denser {\bf k}-mesh used
in the present electronic structure calculation.

\subsection{YBa$_2$Cu$_4$O$_8$}
\begin{figure}
\centering
\vspace*{-0.75in}
\includegraphics[width=1.0\linewidth,trim=1.05in 0in 0.8in 0in, clip=true]{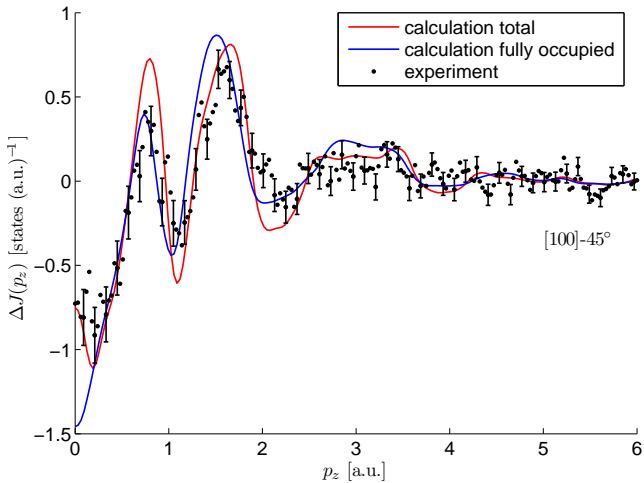}
\vspace*{-2.4in}
\caption{Directional differences of experimental (filled circles) and calculated Compton profiles for all bands
(red line) and only the fully occupied bands (blue line) of YBa$_2$Cu$_4$O$_8$, for scattering vectors along
[100], and the direction 45$^{\circ}$ from [100], rotated about the $c^{*}$-axis.}
\label{ybco}
\end{figure}

Having demonstrated the accuracy of our method in calculating the EMD of a variety of simple structures,
we now apply it to the more electronically and structurally complex cuprate superconductor, YBa$_2$Cu$_4$O$_8$.
The electronic structure of YBa$_2$Cu$_4$O$_8$ was calculated using the crystal structure of Ref. \cite{LIGHTFOOT1991}
with 891 {\bf k}-points in the IBZ, using the PBE-GGA exchange-correlation functional. The resulting band
structure and Fermi surface (not shown) are essentially the same as those calculated previously
\cite{YU1991,AMB1991}. The calculated anisotropy between Compton profiles with scattering vectors along the
[100] direction and a direction 45$^{\circ}$ from [100], rotated about the $c^{*}$-axis is shown in
Fig. \ref{ybco} compared to our experimental data measured on beamline BL08W at the SPring-8 synchrotron, Japan,
with a resolution of 0.11 a.u.. The calculation reproduces the experimental anisotropy well, showing that the
overall description of the electronic structure is satisfactory. For comparison, we also show
the anisotropy when only the fully occupied bands are included in the EMD calculation, such that the difference
between the two anisotropies is entirely due to the FS. It is clear that the calculated anisotropy is
dominated by the contribution from fully occupied bands, and the large oscillatory structure reflects the (projected) wavefunction anisotropy of these bands.

\section{Conclusion}
We have developed a new method for calculating electron momentum densities and Compton profiles from
\textit{ab initio} calculations. The use of the linear tetrahedron method in conjunction with optimally-shaped
tetrahedra allows for accurate momentum density and Compton profile calculations, in any given crystal
structure. This method produces results in excellent agreement with Compton scattering experiments and calculations
produced by other techniques.

\section{Acknowledgements}
The YBa$_2$Cu$_4$O$_8$ experiment was performed with the approval of the Japan Synchrotron Radiation
Institute (JASRI, proposal no. 2011A1483) on single crystals grown by Z. Bukowski. We acknowledge financial
support from the UK EPSRC. This work was carried out using the computational facilities of the Advanced
Computing Research Centre, University of Bristol (\verb+http://www.bris.ac.uk/acrc/+).

\bibliography{daveebib}

\end{document}